\documentclass[amssymb,amsmath,twocolumn,showpacs,aps,pra,longbibliography,superscriptaddress]{revtex4-1}
\usepackage{amsmath}
\usepackage{amsfonts}
\usepackage{amssymb}
\usepackage{dcolumn}
\usepackage{bm}
\usepackage[dvips]{graphicx}
\usepackage{epsfig}

\begin{document}

\title{Proliferation of effective interactions: decoherence-induced equilibration in a closed many-body system}

\author{Pablo R. Zangara}
\affiliation{Instituto de F\'{i}sica Enrique Gaviola (IFEG), CONICET-UNC and Facultad
de Matem\'{a}tica, Astronom\'{i}a y F\'{i}sica, Universidad Nacional
de C\'{o}rdoba, 5000, C\'{o}rdoba, Argentina}
\author{Denise Bendersky}
\affiliation{Instituto de F\'{i}sica Enrique Gaviola (IFEG), CONICET-UNC and Facultad
de Matem\'{a}tica, Astronom\'{i}a y F\'{i}sica, Universidad Nacional
de C\'{o}rdoba, 5000, C\'{o}rdoba, Argentina}
\author{Horacio M. Pastawski}
\email{horacio@famaf.unc.edu.ar}
\affiliation{Instituto de F\'{i}sica Enrique Gaviola (IFEG), CONICET-UNC and Facultad
de Matem\'{a}tica, Astronom\'{i}a y F\'{i}sica, Universidad Nacional
de C\'{o}rdoba, 5000, C\'{o}rdoba, Argentina}

\begin{abstract}
We address the question on how weak perturbations, that are quite
ineffective in small many-body systems, can lead to decoherence and hence to
irreversibility when they proliferate as the system size increases. This
question is at the heart of solid state NMR. There, an initially local
polarization spreads all over due to spin-spin interactions that
conserve the total spin projection, leading to an equilibration of the
polarization. In principle, this quantum dynamics can be reversed by changing the sign of
the Hamiltonian. However, the reversal is usually perturbed by non
reversible interactions that act as a decoherence source. The fraction of the local excitation recovered defines the
Loschmidt echo (LE), here evaluated in a series of closed $N$ spin
systems with all-to-all interactions. The most remarkable regime of the LE
decay occurs when the perturbation induces proliferated effective
interactions. We show that if this perturbation exceeds some lower bound, the decay is ruled by
an effective Fermi golden rule (FGR). Such a lower bound shrinks as $%
N $ increases, becoming the leading mechanism for LE decay in the
thermodynamic limit. Once the polarization stayed equilibrated longer 
than the FGR time, it remains equilibrated in spite of the reversal procedure.

\end{abstract}


\maketitle

\section{INTRODUCTION}

Within the usual wisdom it is quite intuitive to accept that a complex
many-body dynamics could lead to a homogeneous spreading of an initially
localized excitation. Such a process, which in the context of spin systems
has long been known as \textit{spin diffusion}, would lead to the system
equilibration. However, this na\"{\i}ve concept soon encounters limitations.
On one hand, P. W. Anderson discovered that certain conditions preclude the
spreading \cite{Anderson1958}. This problem still generates controversy, as the
question on when do closed many-body quantum systems equilibrate remains
open \cite{polkovnikovRMP,LebowitzQET,Eisert2014,Huse2014}. On the other
hand, even in conditions where the spin diffusion seems
irreversible, nuclear magnetic resonance (NMR) experiments
revealed that the apparently equilibrated state contains correlations
encoding a memory of the initial state \cite{Hahn_atomicMemory}. The pioneer
in this field was E. Hahn. In his spin echo \cite{Hahn1950}, the precession
dynamics of each independent spin is reversed by changing the sign of the
local magnetic fields. In those experiments, the many-spin interaction is
not reversed and consistently it degrades the echo signal in a
characteristic time $T_{2}$. Two decades later, Rhim, Pines and Waugh
exploited the fact that the spin-spin dipolar interaction can also be
reversed \cite{Rhim1971}. This allows for the reversal of a global polarization
state in the form of a Magic Echo. More specific was the development by R.
Ernst and collab. There, a local spin excitation diffuses through a lattice
much as an ink-drop diffuses in a pond. A pulse sequence produces its
refocusing, followed by the local detection as a Polarization Echo \cite%
{Ernst1992}. Again, the attenuation of the observed echo can be tentatively
attributed to the non-inverted terms in the Hamiltonian, as well as
imperfections in the pulse sequence and interactions with some environment.

In a follow up of those experiments, a quest on quantifying the sources that
degrade the echo signal in crystalline samples was initiated \cite%
{patricia98,usaj-physicaA,MolPhys}. As the sources of irreversibility can be
progressively reduced, one might think that there are no limits in the
experimental improvement of the echo. Nevertheless, in systems where a local excitation equilibrates \cite%
{mesoECO-PRL1995,mesoECO-exp}, experiments
show that even weak perturbations are highly effective in
producing the echo degradation. Moreover, there are cases where the time
scale of the decay is intrinsic to the reversed dynamics, i.e. a perturbation independent decay (PID) \cite%
{usaj-physicaA,MolPhys}. If confirmed, this observation could have deep
implications for the degree of controllability of quantum devices as it
evidences a fragility of quantum dynamics towards minuscule perturbations.
In fact, the sensitivity to perturbations or fragility of quantum systems 
\cite{Jacquod2002,Zurek2002,Bendersky2013} is a major problem that transversally affects
several fields, e.g. chaos in quantum computers \cite%
{Shepelyansky2000,Flambaum2000a}, NMR quantum information processing \cite%
{karina09,*Claudia2014,Boutis2012,Cappellaro2013}, quantum criticality \cite%
{Zanardi2006,Silva2008} and, more recently, quantum control theory \cite%
{Calarco2014}.

In order to capture the essentials of the described experiments, the
Loschmidt Echo (LE) is defined as the revival that occurs when an imperfect
time-reversal procedure is implemented \cite{prosen,Jacquod,*Jacquod2006,scholarpedia}. If the unperturbed evolution is given by a classically
chaotic Hamiltonian, there exists a regime in which the decay rate of the
LE corresponds to the classical Lyapunov exponent \cite{jalpa,PazZurek2003}%
. Such a particular PID holds for a semiclassical initial state built from a
dense spectrum and a perturbation above certain threshold. Under weaker perturbations, 
the LE decay depends on their strength following a
Fermi golden rule (FGR) \cite{Jacquod-Beenakker-2001}. Additionally, the LE
semiclassical expansion showed that the PID regime results from the phase
fluctuations along the unperturbed classical trajectories \cite{jalpa}. This
represents a first identification of \textit{irreversibility}, as measured
by the LE, with \textit{decoherence}.

In addressing an actual many-spin dynamics the situation is less clear. On
one side, there is no classical Hamiltonian that serves as reference. On the
other side, the numerical evaluation of the LE in a weakly perturbed \textit{%
finite} spin system, could not justify the experimental observations \cite%
{patricia98}. Since experiments involve almost \textit{infinitely} large systems,
one is left with the question of whether the mechanisms of LE decay and
system equilibration could eventually emerge from a progressive increase in
the system size towards the thermodynamic limit (TL). Here, we tackle such a
question by considering extensive calculations with $N=12,14$ and $16$
interacting spins, whose dynamics involve the complete 2$^{N}$-dimensional
Hilbert space \cite{Dente2013CPC}. We adopt a model with all-to-all dipolar
interactions which allows for small statistical fluctuations facilitating the analysis of
the TL. As in the case of the polarization echo \cite{Ernst1992,patricia98},
our initial state is given by a local excitation in a single spin, and the
detection is also a local measurement of the polarization \cite%
{mesoECO-PRL1995,Zangara2012,Zangara2013PRB}.

We will show that, in the presence of a small Hamiltonian perturbation, the
decay of the LE follows a FGR, much as if the system were interacting with a
continuum. This indicates that the system itself would indeed behave as its
own environment. In addition, we observe that the excitation remains
homogeneously distributed in spite of the time reversal. In other words, the
equilibration produced by the unperturbed Hamiltonian becomes fully
irreversible in the presence of an arbitrary small perturbation. The physical
mechanism responsible for the mentioned FGR corresponds to a \textit{proliferation}
of two- and four-body effective interactions mediated by virtual processes.
Remarkably, we show that the realm of this description is wider as the
system size increases. Such an observation hints that, in the TL, the
proliferation of effective interactions is the sought mechanism that rules
irreversibility.

The paper is organized as follows. In Sec. \ref{Sec_model} we describe the
many-spin model employed to simulate an ideal NMR experiment. Sec. \ref%
{Subsec_autocorrelation} encloses the LE formulation as an autocorrelation
function. In Sec. \ref{Subsec_standardFGR} we discuss the standard FGR
description of the LE. In Sec. \ref{SubSec_effectiveFGR} we introduce the
effective interactions and we use them to evaluate an effective FGR. In
Sec. \ref{Section_numerics} we show the results obtained for the numerical evaluation of the LE time
dependence, including the time-scales and the asymptotic behavior as a function of the
perturbation strength and the system size. Concluding remarks are made in
Sec. \ref{Section_conclusions}.

\section{SPIN MODEL FOR MANY-BODY DYNAMICS\label{Sec_model}}

As in the experimental systems, we consider $N$ spin $1/2$ particles, whose state at $t=0$ is given by the
density matrix:

\begin{equation}
\hat{\rho}_{0}=\frac{1}{2^{N}}(\mathbf{\hat{I}}+2\hat{S}_{1}^{z}).
\label{inistate}
\end{equation}%
Here, $\hat{\rho}_{0}$ stands for a local excitation as $tr[\hat{S}%
_{1}^{z}\hat{\rho}_{0}]=\frac{1}{2}$ and $tr[\hat{S}_{i}^{z}\hat{\rho}%
_{0}]=0 $ $\forall i\neq 1$. The initial polarization is oriented along
the laboratory frame, where the overwhelming Zeeman field of a
superconducting magnet splits the states according to their total spin
projection. Thus, even though the spins would interact through the complete
dipole-dipole interaction, the evolution is ruled by the truncated dipolar
Hamiltonian \cite{Slichter-Book},

\begin{eqnarray}
\hat{H}_{dip} &=&\sum_{i,j}^{N}J_{ij}^{dip}(N)\left[ 2\hat{S}_{i}^{z}\hat{S}%
_{j}^{z}-\left( \hat{S}_{i}^{x}\hat{S}_{j}^{x}+\hat{S}_{i}^{y}\hat{S}%
_{j}^{y}\right) \right]   \label{Hdipolar} \\
&=&\sum_{i,j}^{N}J_{ij}^{dip}(N)\left[ 2\hat{S}_{i}^{z}\hat{S}_{j}^{z}-\frac{%
1}{2}\left( \hat{S}_{i}^{+}\hat{S}_{j}^{-}+\hat{S}_{i}^{-}\hat{S}%
_{j}^{+}\right) \right].
\end{eqnarray}%
This interaction conserves spin projection and is called 
\textit{secular}. Indeed, the symmetry $[ \hat{H}_{dip},\sum_{i=1}^{N}%
\hat{S}_{i}^{z}] =0$ provides the relevant structure of subspaces
given by specific $z$-projections:\ $\nu =\sum_{i=1}^{N}S_{i}^{z}=\frac{N}{2}%
,(\frac{N}{2}-1),...,-\frac{N}{2}$. In a system of $N$ spins, there are $N+1$
subspaces of definite $\nu $, and therefore dynamics induced by $\hat{H}%
_{dip}$ would be strictly confined to each of them.

We choose the coupling strength $J_{ij}^{dip}(N)$ corresponding to an
infinite range or\ all-to-all interaction model, 
\begin{equation}
J_{ij}^{dip}(N)=J_{ji}^{dip}(N)=\left( 1+\chi \right) \times
(-1)^{k}\times \frac{J_{0}}{\sqrt{N}}.  \label{Jdip}
\end{equation}%
Here $\chi$ is a random number taken from a uniform distribution in $%
[-0.1,0.1]$ that ensures the lifting of degeneracies while keeping the
fluctuations of the second moment small. Since the sign of the dipolar
interactions in a crystal depends on the spatial orientation of the
inter-spin vector, we take $k$ as a random number from a binary distribution 
$\{0,1\}$. The price to be paid for an all-to-all network is the absence of
the dynamically hierarchical structure of the experimental systems.

The factor $1/\sqrt{N}$ ensures that the \textit{local} second moment
of the dipolar interaction $\sigma _{dip}^{2}$ remains constant as $N$
changes:

\begin{equation}
\sigma _{dip}^{2}\simeq \sigma _{i}^{2}=\sum_{j(\neq i)}^{N}\left( \frac{%
J_{ij}^{dip}(N)}{2}\right) ^{2}\simeq \frac{J_{0}^{2}}{4}.
\end{equation}%
Therefore, in spite of different cluster sizes, $\hbar /\sqrt{\sigma
_{dip}^{2}}$\ recovers the characteristic spin-spin interaction time $T_{2}$
and consequently $J_{0}$ provides the natural energy unit.

A forward evolution ruled by many-body interactions according to\ $\hat{H}%
_{dip}$ can be experimentally reversed by an appropriate pulse sequence, as
reported in Ref. \cite{MolPhys}. In order to perform the inversion $\hat{%
H}_{dip}\rightarrow -\hat{H}_{dip}$, the full spin state has to be tumbled
down along the direction of a radiofrequency (rf) field that is turned on
immediately afterwards. The rf field rotates perpendicularly to the magnet
one and hence it provides the rotating frame. We redefine
the $z$-direction in such a frame, and thus the rf irradiation yields a Zeeman Hamiltonian:

\begin{equation}
\hat{H}_{Z}=\sum_{i=1}^{N}\hbar \omega _{1}\hat{S}_{i}^{z}.  \label{Hfield}
\end{equation}%
Notice that $\hat{H}_{Z}$ creates finite energy gaps of magnitude $\hbar
\omega _{1}$ which separate the subspaces, but they are not as much effective as
the \textquotedblleft infinite\textquotedblright\ splittings generated by the
magnet (laboratory frame). As a consequence, the Hamiltonian terms that do
not conserve polarization, called \textit{non-secular}, become relevant. The
sign of the non-secular contribution cannot be changed experimentally.
Then, they constitute the perturbation $\hat{\Sigma}$, here embodied by a
double quantum (DQ) Hamiltonian 
\begin{eqnarray}
\hat{\Sigma} = \hat{H}_{dq}&=&\sum_{i,j}^{N}J_{ij}^{dq}(N)\left[ \hat{S}%
_{i}^{x}\hat{S}_{j}^{x}-\hat{S}_{i}^{y}\hat{S}_{j}^{y}\right]  \label{H22} \\
&=&\sum_{i,j}^{N}J_{ij}^{dq}(N)\left[ \hat{S}_{i}^{+}\hat{S}_{j}^{+}+\hat{S}%
_{i}^{-}\hat{S}_{j}^{-}\right].
\end{eqnarray}%
Here, the coupling strength $J_{ij}^{dq}(N)$ satisfies an analogue
definition as in Eq. (\ref{Jdip}). Notice that $[ \hat{H}%
_{dq},\sum_{i=1}^{N}\hat{S}_{i}^{z}] \neq 0$ since $\hat{H}_{dq}$
mixes subspaces whose projections differ in $\delta \nu =\pm 2$ \cite{Pines1985,Pines1987}.
Experimentally, these inter-subspace transitions are partially suppressed by
increasing the rf power, i.e. $\hbar \omega _{1}$\cite%
{UsajTesis,usaj-physicaA}.

\begin{figure}
    \centering
     \includegraphics[width=0.45\textwidth]{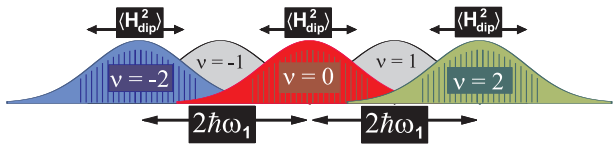}\\
      \caption{Color online. Pictorial representation of the
LDOS of $\hat{H}_{dip}+\hat{H}_{Z}$ respective to the state given in Eq. 
(\ref{inistate}).}%
    \label{Fig_esquema}%
\end{figure}

The information contained in the time domain, embodied in the experimental $%
T_{\mathbf{2}}$ time-scale,\textbf{\ }can be complemented with the spectral
picture given by the Local Density of States (LDOS). This last shows how a
particular state distributes among the eigenstates of a given Hamiltonian.
There is an extensive recent literature recognizing the LDOS as an indicator
for the onset of chaos \cite{Santos2012a,*Santos2012b,Wisniacki2013},
relaxation time-scales \cite{Dente2008,*Dente2011,Wisniacki2009,*Wisniacki2010,Santos2014a,*Santos2014b,*Santos2014c}
and the size of the fluctuations around the steady state \cite%
{Zangara2013PRE}. Even though our evaluation of the dynamics does not rely
on diagonalization \cite{Dente2013CPC}, one can infer the shape of the
unperturbed LDOS ($\hat{H}_{dip}+\hat{H}_{Z}$) respective to the initial
state defined in Eq. (\ref{inistate}). See Fig. \ref{Fig_esquema}. When $%
\hbar \omega _{1}=0$, the subspaces of spin projection are basically
degenerate, and the unperturbed LDOS is a single Gaussian of width $\langle H_{dip}^{2}\rangle =\sum_{i}\sigma _{i}^{2}/4\simeq N\sigma_{dip}^{2}/4$, i.e. the \textit{global} second moment of $\hat{H}_{dip}$. If $\hbar \omega
_{1}\gtrsim \sqrt{\langle H_{dip}^{2}\rangle}  $, the subspaces' LDOS separate from each other.
A subspace with spin projection $\nu $ has mean energy $E_{\nu }\simeq \nu
\hbar \omega _{1}$, and therefore the unperturbed LDOS within each subspace
is:%
\begin{equation}
P_{\nu }(\varepsilon )\simeq \frac{1}{\sqrt{2\pi \langle H_{dip}^{2}\rangle}}\exp \left[ -%
\frac{(\varepsilon -E_{\nu })^{2}}{2\langle H_{dip}^{2}\rangle}\right] .  \label{LDOS}
\end{equation}%
The time domain can be explicitly recovered from the Fourier transform of $%
P_{\nu }(\varepsilon )$\cite{Santos2014a,*Santos2014b,*Santos2014c}.

\section{THE LOSCHMIDT ECHO}

\subsection{The autocorrelation function\label{Subsec_autocorrelation}}

In order to simulate an ideal LE procedure, we assume that forward evolution
occurs under the unperturbed Hamiltonian $\hat{H}_{0}=\hat{H}_{dip}+\hat{H}%
_{Z}$. Even though this evolution would correspond to the laboratory frame,
the addition of the $\hat{H}_{Z}$ term stands for sake of a symmetrical
time-reversal. Besides, as $[ \hat{H}_{dip},\hat{H}_{Z}] =0$,
the inclusion of $\hat{H}_{Z}$ does not introduce any non-trivial dynamics.
At time $t_{R}$, a pulse sequence changes the sign of $\hat{H}_{dip}$, and a \textit{backward} evolution occurs affected by the perturbation.
Hence, the backward dynamics is described by $-\hat{H}_{0}+\hat{\Sigma}=-\hat{%
H}_{dip}-\hat{H}_{Z}+\hat{H}_{dq}$, which in the experiment would correspond
to the rotating frame. The evolution operators for each $t_{R}$-period are $%
\hat{U}_{+}^{{}}(t_{R})=\exp [-\frac{\mathrm{i}}{\hbar }\hat{H}_{0}t_{R}]$
and $\hat{U}_{-}^{{}}(t_{R})=\exp [-\frac{\mathrm{i}}{\hbar }(-\hat{H}_{0}+%
\hat{\Sigma})t_{R}]$ respectively. Then, the LE\ operator,

\begin{equation}
\hat{U}_{LE}^{{}}(2t_{R})=\hat{U}_{-}^{{}}(t_{R})\hat{U}_{+}^{{}}(t_{R}),
\label{ULE}
\end{equation}%
produces an imperfect refocusing at time $2t_{R}$ evaluated as:

\begin{equation}
M_{1,1}(t=2t_{R})=2tr[\hat{S}_{1}^{z}\hat{U}_{LE}^{{}}(t)\hat{\rho}_{0}\hat{U%
}_{LE}^{\dag }(t)].  \label{autocorrelacion1}
\end{equation}%
Since $\hat{S}_{1}^{z}$ is a local (\textquotedblleft
one-body\textquotedblright ) operator, Eq. (\ref{autocorrelacion1}) is
equivalent to the expectation value in a single superposition state \cite%
{Alv-parallelism},

\begin{equation}
M_{1,1}(t)=2\left\langle \Psi _{neq}\right\vert \hat{U}_{LE}^{\dag }(t)\hat{S%
}_{1}^{z}\hat{U}_{LE}^{{}}(t)\left\vert \Psi _{neq}\right\rangle ,
\label{autocorrelacion7}
\end{equation}%
where:

\begin{equation}
\left\vert \Psi _{neq}\right\rangle =\left\vert \uparrow _{1}\right\rangle
\otimes \sum_{r=1}^{2^{N-1}}\frac{1}{\sqrt{2^{N-1}}}e^{\mathrm{i}\varphi
_{r}}\ \left\vert \xi _{r}\right\rangle .  \label{entangled}
\end{equation}%
Here, $\varphi _{r}^{{}}$ is a random phase uniformly distributed in $%
[0,2\pi )$, and $\left\{ \left\vert \xi _{r}\right\rangle \right\} $ stands
for the computational basis states of the Hilbert space corresponding to $%
N-1 $ spins.

\subsection{The standard Fermi golden rule approach \label{Subsec_standardFGR}}

Let us now introduce the regimes of the LE decay, following the dynamical
paradigm from Refs. \cite{jalpa,Jacquod-Beenakker-2001,Lewenkopf2002}. If
the perturbation during the backward evolution is extremely small, the short
time expansion of the LE operator yields a quadratic decay that extends
until recurrences show up. This constitutes the perturbative regime 
\begin{eqnarray}
M_{1,1}(t) &=&2\left\langle \Psi _{neq}\right\vert \hat{U}_{LE}^{\dag }(t)%
\hat{S}_{1}^{z}\hat{U}_{LE}^{{}}(t)\left\vert \Psi _{neq}\right\rangle 
\notag \\
&\simeq &1-\frac{1}{4}\left\langle \Psi _{neq}\right\vert \left[ \hat{\Sigma}%
^{2}-2\hat{\Sigma}\hat{S}_{1}^{z}\hat{\Sigma}\right] \left\vert \Psi
_{neq}\right\rangle t^{2}  \notag \\
&\simeq &1-\left[ t/\tau _{\phi }\right] ^{2}.  \label{LE_shorttime1}
\end{eqnarray}%
Here, $1/\tau _{\phi }$ scales up linearly with the strength of the
perturbation through its local second moment. In fact, $\Sigma _{\alpha
\beta },$ i.e. the matrix elements of $\hat{\Sigma},$ do not exceed the level
spacing $d_{\alpha \beta }$ associated to two
directly connected states (DCS) $\alpha $ and $\beta$.

As the perturbation increases, its second moment exceeds the
typical level spacing among DCS. There, the onset of the Fermi golden
rule occurs. In such a case, the perturbed energies $\widetilde{E}_{\alpha }$
are obtained from the unperturbed $E_{\alpha },$ using second
order perturbative series, which must be evaluated in the TL:

\begin{equation*}
\widetilde{E}_{\alpha }\simeq E_{\alpha }+\lim_{\eta \rightarrow
0^{+}}\lim_{N\rightarrow \infty }\sum_{\beta }\frac{\left\vert \Sigma
_{\alpha \beta }\right\vert ^{2}}{d_{\alpha \beta }+\mathrm{i}\eta }%
=E_{\alpha }+\Delta _{\alpha }-\mathrm{i}\Gamma _{\alpha },
\end{equation*}%
where the real shift $\Delta _{\alpha }$ and the imaginary correction $%
\Gamma _{\alpha }$ are defined as 
\begin{eqnarray}
\Delta _{\alpha } &=&\mathcal{P}\sum_{\beta }\frac{\left\vert \Sigma _{\alpha
\beta }\right\vert ^{2}}{d_{\alpha \beta }},  \label{Delta_alfa} \\
\Gamma _{\alpha } &=&2\pi \sum_{\beta }\left\vert \Sigma _{\alpha \beta
}\right\vert ^{2}\delta (E_{\beta }-E_{\alpha }).  \label{Gama_alfa}
\end{eqnarray}%
Here $\mathcal{P}$ stands for Principal value. In most practical cases,
$\Delta _{\alpha }$ provides a small energy shift that can be neglected.
Notice that the decay introduced by $\Gamma _{\alpha }$ requires the 
\textit{mixing of infinitely many quasidegenerate states}. Additionally,
$\Gamma _{\alpha }$ can be replaced by its local energy-average,

\begin{equation}
\left\langle \Gamma \right\rangle =2\pi \left\langle \Sigma
^{2}\right\rangle /d.  \label{gama_average}
\end{equation}%
Here, $d$ stands for the mean level spacing among the DCS.
Therefore, within the standard FGR approximations, a single LE
operator already contains a decay 
\begin{flalign}
\hat{U}_{LE}^{}(t)&\simeq \sum\limits_{\alpha }e^{\mathrm{i}\widetilde{E}%
_{\alpha }t/\hbar }e^{-\mathrm{i}E_{\alpha }t/\hbar }\left\vert \alpha
\right\rangle \left\langle \alpha \right\vert \simeq \nonumber\\ 
& \qquad \qquad  \simeq \sum\limits_{\alpha
}e^{-\Gamma _{\alpha }t/\hbar }\left\vert \alpha \right\rangle \left\langle
\alpha \right\vert \simeq e^{-\left\langle \Gamma \right\rangle t/\hbar }%
\mathbf{\hat{I}.}  \label{LE_FGRapprox}
\end{flalign}%
This constitutes the standard Random Matrix Theory (RMT) approach to the LE 
\cite{Jacquod-Beenakker-2001,Lewenkopf2002}.

\subsection{From virtual interactions to an effective Fermi golden rule\label{SubSec_effectiveFGR}}

As pointed above, the non-secular DQ perturbation $\hat{\Sigma}$
only mixes states from different Zeeman subspaces. Then, the previous FGR
requirement of mixing quasidegenerate states is not fulfilled. Nevertheless, as hinted by the experiments \cite{usaj-physicaA}, the DQ interaction could produce
effective secular terms of major relevance in the TL. We will now formalize these ideas showing how a small
non-secular perturbation can connect quasidegenerate states through virtual
processes.

Given a specific total spin projection $\nu $, its corresponding subspace $%
\mathcal{S}_{\nu }$ is coupled to the subspaces $\mathcal{S}_{\nu +2}$ and $%
\mathcal{S}_{\nu -2}$ by the DQ interaction. In other words, the
perturbation $\hat{\Sigma}$ produces transitions with $\delta \nu =\pm 2$
that involve an energy difference of $2\hbar \omega _{1}$. However, there
are higher order transitions that avoid the energy mismatch. For instance,
when state $\left\vert \uparrow \downarrow \downarrow \right\rangle $ swaps
to $\left\vert \uparrow \uparrow \uparrow \right\rangle $ and then to $%
\left\vert \downarrow \downarrow \uparrow \right\rangle $ (back to the initial subspace), one gets an the
effective flip-flop between spins $1$ and $3$. This constitutes an
intra-subspace effective coupling of order $(J^{dq})^{2}/(\hbar \omega _{1})$%
. A more sophisticated process occurs when $\left\vert \uparrow \uparrow
\downarrow \downarrow \right\rangle $ swaps to $\left\vert \uparrow \uparrow
\uparrow \uparrow \right\rangle $ and then back to $\left\vert \downarrow
\downarrow \uparrow \uparrow \right\rangle $. It provides for a four-body
effective interaction. Therefore, if the energy gaps $\hbar \omega _{1}$ are
large enough, \textit{inter}-subspace transitions are in fact truncated, but
then \textit{intra}-subspace transitions mediated by satellite subspaces set
in. These lead us to the corresponding effective Hamiltonian,

\begin{equation}
\hat{V}_{eff}\simeq \sum_{k,l}^{N}\sum_{i,j}^{N}\frac{J_{lk}^{dq}J_{ij}^{dq}%
}{8\hbar \omega _{1}}\left( \hat{S}_{l}^{+}\hat{S}_{k}^{+}\hat{S}_{i}^{-}%
\hat{S}_{j}^{-}+\hat{S}_{l}^{-}\hat{S}_{k}^{-}\hat{S}_{i}^{+}\hat{S}%
_{j}^{+}\right) .  \label{virtual8}
\end{equation}%
Such a result finds a formal justification either on a Green's function
approach to the Effective Hamiltonian \cite{ChemPhys2002_eph} or in the
Average Hamiltonian theory \cite{Waugh1968}. It is crucial to notice that $%
\hat{V}_{eff}$ can indeed mix quasidegenerate states within a particular $%
\mathcal{S}_{\nu }$. Furthermore, it can couple states in $\mathcal{S}_{\nu }$
that were not originally coupled by $\hat{H}_{dip}$. In practice, this
means that effective matrix elements do appear in places where the original
raw $\hat{H}_{dip}$ had null entries, leading to a remarkable \textit{%
proliferation} of interactions.

In principle, destructive interferences can take place. For instance, the
transition from state $\left\vert \uparrow \uparrow \downarrow \downarrow
\right\rangle $ to $\left\vert \uparrow \uparrow \uparrow \uparrow
\right\rangle $ and then back to $\left\vert \downarrow \downarrow \uparrow
\uparrow \right\rangle $ would cancel out the transition from $\left\vert
\uparrow \uparrow \downarrow \downarrow \right\rangle $ to $\left\vert
\downarrow \downarrow \downarrow \downarrow \right\rangle $ and then back to 
$\left\vert \downarrow \downarrow \uparrow \uparrow \right\rangle $. Many of
the destructive interferences enabled by a homogeneous all-to-all model,
i.e. $J_{lk}^{dq}=J_{ij}^{dq}$ for any $l,k,i,j$ indexes, are nevertheless
removed by the randomization of parameters $k$ and $\chi$. Other
realistic spin models, in which the strength and sign of the spin-spin
interaction depends on the relative positions of the spins, would not
exhibit such an interference. Based on the same argument, the effective hopping
corrections in Eq. (\ref{virtual8}) generate almost random entries in the
Hamiltonian of each subspace. This proliferation may justify modelling the
dynamics through standard RMT instead of the two-body random ensembles \cite%
{Bohigas1971,Flambaum96}.

The natural step now consists in formulating an effective FGR description as
in the RMT approach introduced in Sec. \ref{Subsec_standardFGR}. Accordingly, we define the global second
moment of the virtual interactions: 
\begin{equation}
\left\langle V_{eff}^{2}\right\rangle =\left\langle \sum_{\beta }\left\vert
\left\langle \beta \right\vert \hat{V}_{eff}\left\vert \alpha \right\rangle
\right\vert ^{2}\right\rangle _{\alpha }=\left\vert a\frac{\left(
J^{dq}\right) ^{2}}{2\hbar \omega _{1}}\right\vert ^{2},
\label{Veff_segundoMomento}
\end{equation}%
where $a$ is a geometrical coefficient that counts the average number of
states connected to a given state $\alpha $. Also, $\left\langle \cdot
\right\rangle _{\alpha }$ denotes the average over all unperturbed
eigenstates $\alpha $. In analogy to Eq. (\ref{gama_average}),%
\begin{equation}
\Gamma _{eff}\sim 2\pi \left\langle V_{eff}^{2}\right\rangle
d_{eff}^{-1}=2\pi \left\vert a\frac{\left( J^{dq}\right) ^{2}}{2\hbar \omega
_{1}}\right\vert ^{2}d_{eff}^{-1},  \label{Gama_FGReff}
\end{equation}%
where $d_{eff}^{-1}$ is the density of DCS by the
virtual interaction. It can be estimated as $d_{eff}\sim bJ^{dip}$ for some
geometrical coefficient $b\ll 1$. Both $a$ and $b$ stand for a subtle
interplay between $N$, the coordination number of the lattice, the selection
rules of the interaction, etc.

In what follows, we present a numerical study of the LE dynamics to show
how it depends on the strength of the effective perturbation $\Sigma
_{eff}=\left( J^{dq}\right) ^{2}/(\hbar \omega _{1})$. One of the purposes consists in 
finding the applicability of the effective FGR.

\section{LOSCHMIDT ECHO NUMERICAL EVALUATION\label{Section_numerics}}

\begin{figure}
    \centering
     \includegraphics[width=0.48\textwidth]{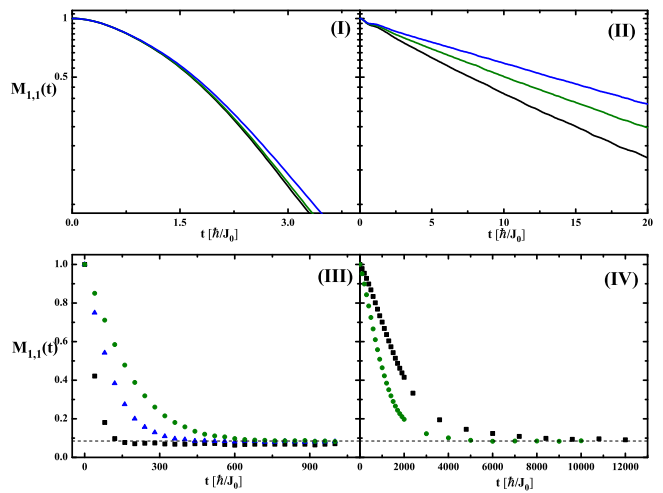}\\
      \caption{Color online. LE time
dependence, $N=14$. The magnitudes of $(J^{dq})^{2}/(\hbar \protect\omega %
_{1})$ are, from top to bottom, (I): $0.67J_{0}$, $1.35J_{0},$ $\infty $;
(II): $0.19J_{0}$, $0.21J_{0}$, $0.23J_{0}$;\ (III): $0.071J_{0}$, $%
0.048J_{0}$, $0.038J_{0}$; (IV): $0.013J_{0}$, $0.009J_{0}$. Plots (I) and
(II) are in log-linear scale, while plots (III) and (IV) are in linear scale.}%
    \label{Fig_LE1}%
\end{figure}

Fig. \ref{Fig_LE1} shows the typical LE dynamics for different perturbation
strengths $\Sigma _{eff}$. In particular, Figs. \ref{Fig_LE1} (I) and (II)
show a Gaussian to exponential transition as $\Sigma _{eff}$ decreases. A
similar transition has been reported for the Survival Probability of
specific many-body states \cite{Flambaum2001,Santos2012a,*Santos2012b}.

\begin{figure}
    \centering
     \includegraphics[width=0.40\textwidth]{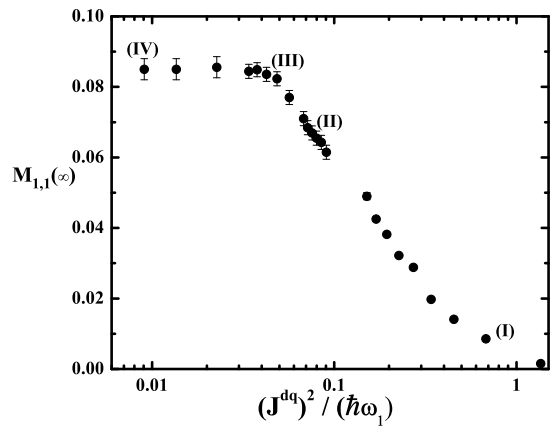}\\
      \caption{LE asymptote $%
M_{1,1}(t\rightarrow \infty )$ as a function of $(J^{dq})^{2}/(\hbar \protect%
\omega _{1})$ (in units of $J_{0}$). The labels I, II, III and IV correspond
to the representative cases shown in Fig. \protect\ref{Fig_LE1}. Data set
corresponds to $N=14$.}%
    \label{Fig_LEasintotas}%
\end{figure}

Figs. \ref{Fig_LE1} (III) and (IV) show an asymptotic plateau for $%
M_{1,1}(t) $ that sets in when the perturbation is small enough (i.e. large $%
\hbar \omega _{1}$). In order to quantify such an observation, we plot in Fig. %
\ref{Fig_LEasintotas} the LE asymptotic plateau $M_{1,1}(t\rightarrow \infty
)$ as a function of $\Sigma _{eff}$. Below a perturbation threshold, say $%
\Sigma _{eff}\lesssim 0.05J_{0}$ in Fig. \ref{Fig_LEasintotas}, the LE
equilibrates slightly above $1/N$. The asymptotic equidistribution $1/N$ becomes very precise
for $N$ above 18 (data not shown). It is important to notice that this
equilibration goes beyond the raw one that occurs in the forward evolution,

\begin{equation}
2\left\langle \Psi _{neq}\right\vert \hat{U}_{+}^{\dag }(t)\hat{S}_{1}^{z}%
\hat{U}_{+}^{{}}(t)\left\vert \Psi _{neq}\right\rangle \underset{%
(t\rightarrow \infty )}{\longrightarrow }1/N.  \label{asintotaForward}
\end{equation}%
Indeed, a perfect reversal of $\hat{U}_{+}^{{}}(t)$\ would unravel the
equilibration stated in Eq. (\ref{asintotaForward}). Nevertheless, the fact
that $M_{1,1}(t\rightarrow \infty )$\ still keeps $\sim 1/N$\ means that the
perturbation stabilizes the spreading of the spin polarization, turning such a
process into an irreversible phenomenon. In addition, one should notice that
the final state conserves the total spin projection despite of the
non-conserving nature of the DQ perturbation. In fact, this evidences the
relevance of the effective interactions discussed in Sec. \ref{SubSec_effectiveFGR}, since they
provide a LE decay mechanism without compromising the conservation of spin projection.

\begin{figure}
    \centering
     \includegraphics[width=0.45\textwidth]{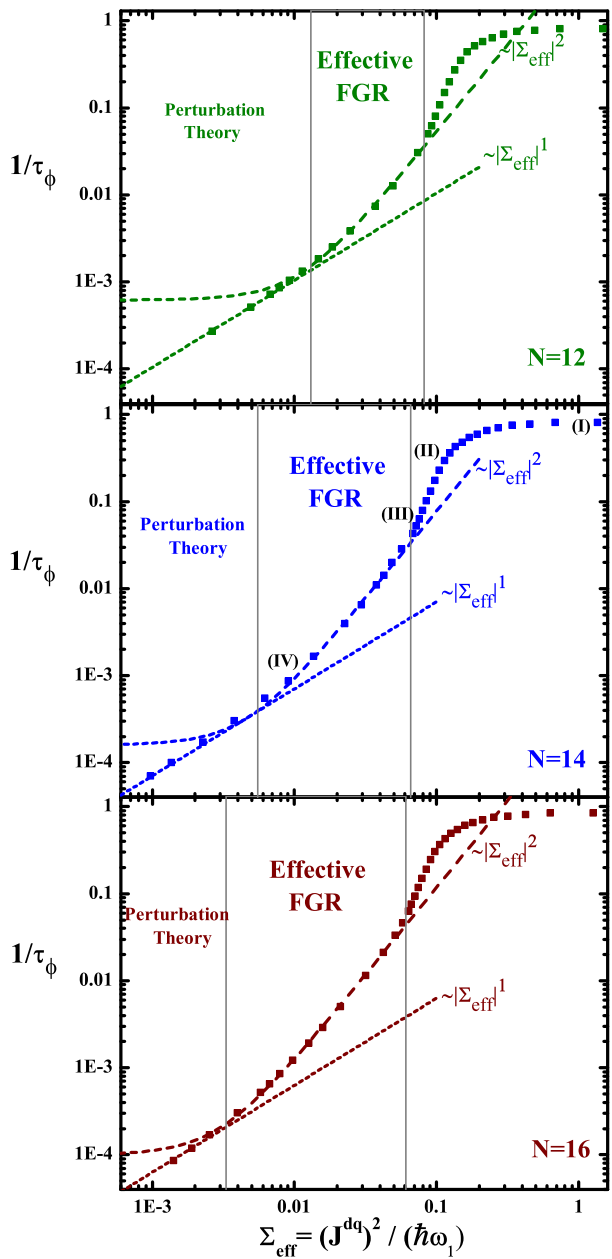}\\
      \caption{Color online. LE decay rates $1/%
\protect\tau _{\protect\phi }$ (log scale, in units of $J_{0}/\hbar $) as a function of
the effective perturbation $\Sigma _{eff}=(J^{dq})^{2}/(\hbar \protect\omega %
)$ (units of $J_{0}$), for $N=12,14,16$. The labels I, II, III and IV from
Figs. \protect\ref{Fig_LE1} and \protect\ref{Fig_LEasintotas} are included
in the case $N=14$.}%
    \label{Fig_LErates}%
\end{figure}

In order to quantitatively assess the LE decay, we define its characteristic
time $\tau _{\phi }$ as $M_{1,1}(\tau _{\phi })=2/3$. We plot the rates $%
1/\tau _{\phi }$ in Fig. \ref{Fig_LErates} as a function of $\Sigma _{eff}$
for $N=12,14,16$. For each size, we identify the regimes in which the rate
scales linearly and quadratically with $\Sigma _{eff}$. The former case can
be understood as being strictly perturbative, i.e. Eq. (\ref{LE_shorttime1}). The latter is associated to the effective FGR, i.e. Eq. (\ref{Gama_FGReff}), as $1/\tau _{\phi }-1/\tau _{\phi ,N}^{0}\propto $ $\Sigma_{eff}^{2}$ and $1/\tau _{\phi ,N}^{0}\rightarrow 0$ as $N\rightarrow \infty
$. The vanishing $1/\tau _{\phi ,N}^{0}$ in the present all-to-all model,
differs from the rate offsets observed in some hierarchical lattices \cite%
{Zangara2012}. The numerical observation that the effective FGR onset moves
steadily towards weaker perturbations as $N$ increases constitutes the main
result of our paper.

The comparison between Figs. \ref{Fig_LEasintotas} and \ref{Fig_LErates} for
the $N=14$ case evidences that the regime where the effective FGR is valid
coincides with the $\sim 1/N$ equilibration of the spin polarization. This
contrasts with the non-ergodic behavior expected for the perturbative
regime. In terms of time scales, given an arbitrarily small perturbation
characterized by its corresponding FGR time $\tau _{\phi }$, if the forward evolution $\hat{U%
}_{+}^{{}}(t)$ occurs for a time $t\gg \tau _{\phi }$, then the
equilibration in Eq. (\ref{asintotaForward}) becomes irreversible for any
practical purpose.

As compared with equilibration, thermalization constitutes a much more
specific process \cite{Eisert2014,Huse2014}. We can identify the initial
condition of our system, i.e. Eq. (\ref{inistate}), as an infinite temperature equilibrium state plus
an excitation. Given the impossibility to revert the dynamics, the
correlations are useless and the system ends up cooled down to a finite
temperature state. We do not elaborate further along this line as it would
not contribute to our central discussion.

\section{CONCLUSIONS\label{Section_conclusions}}

We have computed the LE, here defined as the local polarization recovered after a
perturbed time reversal procedure, showing a wealth of dynamical regimes.
The dynamics of clusters of interacting
spins has been evaluated employing their complete Hilbert space. In order to analyze the
emergence of the TL as $N$ increases up to $16$ spins, we have adopted an
all-to-all interaction model. Forward dynamics is generated by a reversible
truncated dipolar Hamiltonian $\hat{H}_{0}$ that provides a natural
decomposition of the Hilbert space into subspaces of definite spin
projection. As in the original experiments, a non-reversible perturbation $%
\hat{\Sigma}$ couples subspaces which are separated by controllable energy
gaps.

We address a regime in which the perturbation induces two- and four-body
effective interactions that can mix quasidegenerate states. These states were not directly
coupled by the dipolar Hamiltonian. Moreover, since the effective interactions have 
fewer restrictions to the selection rules, they proliferate
within each subspace. In such a regime, the LE decay is characterized by an
effective FGR whose realm of validity widens towards weaker perturbations as 
$N$ increases. The analysis of this lower bound follows a specific
sequence for the two limits: first $N$ $\rightarrow \infty $ and then $%
\Vert \hat{\Sigma}\Vert \rightarrow 0^{+}$. Then, in the TL, even
a slight perturbation yields a LE decay ruled by an effective FGR, which is enhanced
by the mechanism of proliferation of effective interactions.

In our model, forward many-spin dynamics can already yield an asymptotic
equidistribution of the polarization. Remarkably, we observe that the
excitation remains homogeneously distributed in spite of the time reversal.
Therefore, while the equilibrated state indeed contains correlations that
encode a full memory of the initial state, such correlations are useless
in the presence of arbitrarily small perturbations. These would render the time
reversal of the Hamiltonian completely ineffectual.

\section{ACKNOWLEDGEMENTS}

This work benefited from discussions with L. Buljubasich, A. K. Chattah, C.
Cormick, L. Foa Torres, F. Pastawski, C. Pineda, C. S\'{a}nchez, L. F.
Santos, and D. A. Wisniacki. We are grateful to C. Bederi\'{a}n for the
technical support on GPU hardware. We acknowledge support from CONICET,
ANPCyT, SeCyT-UNC and MinCyT-Cor. This work used Mendieta Cluster from CCAD
at UNC, that is part of SNCAD-MinCyT, Argentina.


%

\end{document}